\documentclass[prd,showpacs,superscriptaddress,nofootinbib,floatfix,showkeys,10pt]{revtex4}
\usepackage{graphicx}
\usepackage{amsmath}
\usepackage{bm}
\usepackage{yhmath}
\usepackage{mathtools}
\usepackage{wasysym}
\usepackage[colorlinks,citecolor=violet,urlcolor=violet,linkcolor=blue]{hyperref}
\usepackage{subfigure}
\usepackage{color}
\usepackage{cases}
\usepackage{subfigure}
\usepackage{times}
\usepackage{dcolumn,booktabs,bm}
\usepackage{slashed}
\usepackage{amsfonts,amssymb,stmaryrd,latexsym,amsmath}
\usepackage{textcomp}
\usepackage{multirow}
\usepackage{cancel}
\usepackage{array}
\usepackage{orcidlink}
\usepackage{longtable}
\usepackage[utf8]{inputenc}
\usepackage[T1]{fontenc}
\usepackage[utf8]{inputenc}
\usepackage{cprotect}
\usepackage{fvextra}
\usepackage{textcomp}

\begin{document}
\title{Thermodynamics of Rotating Lifshitz Dilaton Black Brane in Quartic Quasitopological gravity}

\author{M. Ghanaatian\,\orcidlink{0000-0002-7853-6767}}\email{mghanaatian@jahromu.ac.ir}\thanks{Corresponding Author}
\affiliation{Department of Physics, Jahrom University, Jahrom, P. O. Box 74137-66171, Iran}

\author{F. Esmaeili\,}
\affiliation{Department of  Physics, Payame Noor University (PNU), P. O. Box 19395-3697 Tehran, Iran}

\author{A.~Bazrafshan\,}
\affiliation{Department of Physics, Jahrom University, Jahrom, P. O. Box 74137-66171, Iran}

\author{Gh. Forozani\,}
\affiliation{Department of  Physics, Payame Noor University (PNU), P. O. Box 19395-3697 Tehran, Iran}

% In the plane
% at the threshold
% on the sheets

\begin{abstract}
In this paper, we obtain the rotating Lifshitz dilaton black brane solutions in the presence of the quartic quasitopological gravity and then probe the related thermodynamics. At first, we obtain the field equations form which a total constant along the radial coordinate $r$ is deduced. Since we cannot solve the solutions exactly, so we investigate their asymptotic behaviors at the horizon and at the infinity. We attain the conserved and thermodynamic quantities such as temperature, angular velocity, entropy, the energy and the angular momentum densities of the rotating quartic quasitopological Lifshitz dilaton black brane. By evaluating the total constant at the horizon and the infinity, we can make a relation between the thermodynamic quantities and so get to a Smarr-type formula. We demonstrate that the thermodynamic quantities of this rotating black brane obey the first law of the thermodynamics. We also study the thermal stability of the rotating quartic quasitopological Lifshitz dilaton black brane and it is not thermally stable. 
\end{abstract}


\maketitle

\section{Introduction}
AdS/CFT correspondence is a useful tool to relate the strongly-coupled systems in quantum field theory to the weakly-coupled gravitational ones in anti-de Sitter (AdS) spacetime \cite{Malda,Witten,Gubser,Aharony}. Based on this duality, some unconventional condensed matter systems are described at their critical temperatures by the nonrelativistic version of the AdS/CFT \cite{Balasu,Hart,Herz,McGr0}. The such quantum field theories have 
an anisotropic scale symmetry in which the scale invariance is established but time and space scale differently. Lifshitz spacetimes can be dual candidates for the such nonrelativistic systems with the description \cite{Kachru} 
\begin{eqnarray}\label{metric1}
ds^2=-\frac{r^{2z}}{l^{2z}}dt^2+\frac{l^2}{r^2}dr^2+r^2d\vec{X}^2,
\end{eqnarray}
where $z>1$ and $l$ are respectively the dynamical critical exponent and the AdS radius. This metric is not conformally
invariant and it shows an anisotropic scale invariance of the type
\begin{eqnarray}
t\rightarrow \zeta^{z}t,\,\,\,\vec{X}\rightarrow \zeta \vec{X}.
\end{eqnarray}
In the Lifshitz metric, a homogeneity may be seen with time and space translation invariance, spatial rotational symmetry, spatial parity and time reversal invariance. Nonrelativistic
conformal field theories depicting multicritical points in certain magnetic materials and liquid crystals can be duals for the Lifshitz spacetimes \cite{Kachru}. Some ones have studied the quantities such as viscosity and conductivity for the systems with Lifshitz symmetry in Refs. \cite{Pang} and \cite{Sin0,Lemos22,Qian,Roychowdhury}. Lifshitz black holes with exact and numeric solutions have been investigated in Refs. \cite{Aze,Danie,Bala,Ayon,Bert0,Mann0,Bert1,Pour1,Bryn,kord2}.\\
Based on the gauge/gravity duality, there is a relation between the central charges on the quantum side and the coupling parameters
on the gravitational side. Einstein gravity does not have enough free parameters to relate them to the central charges in the CFTs. Also, Einstein gravity cannot lead to a Lifshitz solution, unless one considers a modified gravity with higher-order curvature tensors or a matter source. A generalization to include higher curvature corrections terms to Einstein gravity and thermodynamics of asymptotically Lifshitz black holes have been done \cite{Bravo}. Lifshitz black holes with higher curvature gravities are in Refs. \cite{Garb,Hohm,Mann3,Mann1,Asnafi,Mann2,Bravo,Bazr1}. In Refs. \cite{Taylor,Bert,Maeda,Lee,Tarrio,Gonz,Mirza1}, Lifshitz black holes with matter sources such as dilaton and Brans-Dicke scalars have been considered. In this regard, thermal behaviors of uncharged \cite{Bert} and linearly charged \cite{Tarrio} Lifshitz black holes in the context of the dilaton gravity have been also probed.  \\
Quasitopological gravity is one the most important modified gravities that has been studied in Refs. \cite{Mann2,Bazr0,Bazr1,Bazr3,Bazr2,Mirza2}. It can provide more free parameters that have been investigated holographically in Ref. \cite{Myers}. At first glance, it seems that the quasitopological gravity is similar to the LoveLock theory \cite{Lovelock}. However the quasitopological gravity has some priorities over the LoveLock one. Lovelock theory is the most general one of the gravitation that is quasilinear in the second order derivatives of the metric and in each dimension, it has no higher order derivatives for a general spacetime. In the presence of the $k$-th order Lovelock theory, one can obtain the field equations only for $k\leq\big[\frac{n}{2}\big]$, (where $n$ counts the space dimensions) and it is impossible to obtain the field equations in higher dimensions. This is while that for a $k$-th powered curvature in the quasitopological gravity, the action is valid for all dimensions larger than four except for the case $n=2k-1$. This is because of the quasitopological terms of this gravity. For example, as the fourth order Lovelock gravity is
effective only for the dimension nine or above, the quasitopological gravity with the fourth order curvature tensors is effective for all dimensions, $n\geq 4$, except for $n=7$. In this paper, we are eager to probe the quasitopological gravity with the fourth order curvatures named as the quartic quasitopological gravity. Based on the AdS/CFT correspondence, adding a quartic curvature term with a new coupling constant on the gravity side can provide a broader class
of four (and higher)-dimensional CFTs. In the quartic quasitoplogical gravity, the equations of motions are still second order and the black hole solutions are explicit \cite{Bazr0}. Also, it is the largest order of the gravity for which the field equations can be solved analytically, while it is not possible to obtain this largest analytic solution in the Lovelock theory. Another interesting issue that shows the superiority of the quasitopological gravity is causality. The establishment of the causality in the dual CFT can make a constraint on the coupling constants of the gravity theory so that the dual theory does not sustain superluminal signals.
In the quasitopological gravity, the causality is preserved since the constraints coming from the requiring positive energy
fluxes condition are in accordance with the causality ones \cite{Myers}. In Lovelock gravity this matching is confirmed \cite{Yang00} but not in general, especially for the cases having no second order gravitational equations of motion \cite{Hofman}. \\
Now we want to investigate the thermodynamics of the rotating Lifshitz dilaton black brane in the presence of the quartic quasitopological gravity. So in Sec. \eqref{Field}, we define the main structure of the Lifshitz dilaton quartic quasitopological theory and then obtain some constraints requiring the existence of the Lifshitz solutions. Next in Sec. \eqref{eauation}, we obtain the field equations and then obtain a constant along the radial coordinate $r$. We probe the rotating quartic quasitopological black brane solutions in the two near-horizon and infinity regimes. In Sec.\ref{thermo}, we obtain a finite and well-defined action for the rotating quartic quasitopological Lifshitz dilaton black brane. We also obtain the conserved and themodynamic quantities and then probe the thermal stability of the black brane. At last, we have a brief conclusion of the whole paper in Sec. \ref{result}. Some notes are also in the App. \ref{app1}.
%%%%%%%%%%%%%%%%%%%%%%%%%%%%%%%%%%%%%%%%%%%%%%%%%%%%
\section{Basic Formalism}\label{Field}
Our theory starts with the quartic quasitopological gravity in the presence of a gauge field $A_{\mu}$ coupled to a dilaton field $\Phi(r)$. So, the related $(n+1)$-dimensional bulk action is 
\begin{eqnarray}\label{Action}
I_{b}&=&\frac{1}{16\pi}\int_{\mathcal{M}}d^{n+1}x \sqrt{-g}\bigg[\sum_{i=1}^{4}\alpha_{i}\mathcal{L}_{i} -2\Lambda-\frac{4}{n-1}(\partial \Phi)^2 -e^{-4\lambda \Phi/(n-1)}F_{\mu\nu}F^{\mu\nu}\bigg],
\end{eqnarray}
where $\mathcal{L}_{1}$, $\mathcal{L}_{2}$, $\mathcal{L}_{3}$ and  $\mathcal{L}_{4}$ are respectively the Hilbert-Einstein, Gauss-Bonnet, cubic and quartic quasitopological gravities with the bellow definitions
\begin{eqnarray}\label{Lag}
\mathcal{L}_{1}&=&R,\nonumber\\
\mathcal{L}_{2}&=&R_{abcd}R^{abcd}-4R_{ab}R^{ab}+R^2,\nonumber\\
\mathcal{L}_{3}&=&
R_a{{}^c{{}_b{{}^d}}}R_c{{}^e{{}_d{{}^f}}}R_e{{}^a{{}_f{{}^b}}}+\frac{1}{(2n-1)(n-3)}\bigg[\frac{3(3n-5)}{8}R_{abcd}R^{abcd}R-3(n-1)R_{abcd}R^{abc}{{}_e}R^{de}\nonumber\\
&&+3(n+1)R_{abcd}R^{ac}R^{bd}+6(n-1)R_a{{}^b}R_b{{}^c}R_{c}{{}^a}-\frac{3(3n-1)}{2}R_a{{}^b}R_b{{}^a}R +\frac{3(n+1)}{8}R^3\bigg]\nonumber\\
\mathcal{L}_{4}&=& c_{1}R_{abcd}R^{cdef}R^{hg}{{}_{ef}}R_{hg}{{}^{ab}}+c_{2}R_{abcd}R^{abcd}R_{ef}{{}^{ef}}+c_{3}RR_{ab}R^{ac}R_c{{}^b}+c_{4}(R_{abcd}R^{abcd})^2\nonumber\\
&&+c_{5}R_{ab}R^{ac}R_{cd}R^{db}+c_{6}RR_{abcd}R^{ac}R^{db}+c_{7}R_{abcd}R^{ac}R^{be}R^d{{}_e}+c_{8}R_{abcd}R^{acef}R^b{{}_e}R^d{{}_f}\nonumber\\
&&+c_{9}R_{abcd}R^{ac}R_{ef}R^{bedf}+c_{10}R^4+c_{11}R^2 R_{abcd}R^{abcd}+c_{12}R^2 R_{ab}R^{ab}\nonumber\\
&&+c_{13}R_{abcd}R^{abef}R_{ef}{{}^c{{}_g}}R^{dg}+c_{14}R_{abcd}R^{aecf}R_{gehf}R^{gbhd},
\end{eqnarray}
that the coefficients $c_{i}$'s are in the appendix.\eqref{app1}. For simplicity in calculations, we redefine the quasitopological coefficients $\alpha_{i}$'s  by 
\begin{eqnarray}\label{redi}
\alpha_{1}&=&1\,, \nonumber\\
\alpha_{2}&=&\frac{\hat{\alpha}_{2}}{(n-2)(n-3)}\,, \nonumber\\
\alpha_{3}&=&\frac{8(2n-1)\hat{\alpha}_{3}}{(n-2)(n-5)(3n^2-9n+4)}\,,\nonumber\\
\alpha_{4}&=&\frac{\hat{\alpha}_{4}}{n(n-1)(n-3)(n-7)(n-2)^2(n^5-15n^4+72n^3-156n^2+150n-42)}.
\end{eqnarray}
In the action \eqref{Action}, $\Lambda$ and $\lambda$ are respectively the cosmological constant and the coupling strength of the dilaton and the gauge field which obeys the bellow relation 
\begin{eqnarray}\label{pot}
F_{\mu\nu}=\partial_{\mu}A_{\nu}-\partial_{\nu}A_{\mu}.
\end{eqnarray}
We intend to obtain the $(n+1)$-dimensional rotating Lifshitz black brane solutions with a rotation parameter $a$, so the metric is specified by
\begin{eqnarray}\label{metric2}
ds^2=-\frac{r^{2z}}{l^{2z}}f(r)(\Xi dt-ad\phi)^2+\frac{l^2}{r^2 g(r)}dr^2+\frac{r^2}{l^4}(a dt-\Xi l^2 d\phi)^2+r^2\sum_{i=1}^{n-2}d\vec{X}_{i}^2,
\end{eqnarray}
where $\Xi=\sqrt{1+a^2/l^2}$. The static Lifshitz metric \eqref{metric1} and the rotating one \eqref{metric2} are mapped into each other by the local transformation 
\begin{eqnarray}
t\rightarrow\Xi t-a\phi,\,\,\,\,\,\,\,\,\,\,\,\phi\rightarrow\Xi \phi-\frac{a}{l^2}t,
\end{eqnarray}
where for $a=0$, the rotating metric \eqref{metric2} reduces to the static metric \eqref{metric1}. So by choosing $a=0$ or $\Xi=1$, this paper includes another class of the solutions as the static quartic quasitopological Lifshitz dilaton black brane ones. One may think that the static metric \eqref{metric1} and the rotating one \eqref{metric2} are the same but, if the coordinate $\phi$ has a periodic behavior, then the rotating metric can provide a new spacetime. This leads that the two metrics are mapp into each other locally not globally. It should be noted that $d\vec{X}^2$ represents a $(n-1)$-dimensional hypersurface with the constant curvature zero in a volume $V_{n-1}$. We use the ansatz
\begin{eqnarray}\label{ans}
A_{\mu}=q e^{K(r)}(\Xi dt-a d\phi)=\frac{q}{l^z}k(r)(\Xi dt-a d\phi),
\end{eqnarray}
where $q$ is a constant which will be bound later and we have used the transformation 
\begin{eqnarray}\label{trans1}
K(r)&=&\mathrm{ln}\bigg[\frac{k(r)}{l^z}\bigg].
\end{eqnarray}
It is easy to show that the action \eqref{Action} supports the Lifshitz solution
\begin{eqnarray}\label{metric3}
ds^2=-\frac{r^{2z}}{l^{2z}}(\Xi dt-ad\phi)^2+\frac{l^2}{r^2}dr^2+\frac{r^2}{l^4}(a dt-\Xi l^2 d\phi)^2+r^2\sum_{i=1}^{n-2}d\vec{X}_{i}^2,
\end{eqnarray}
if the bellow conditions are satisfied
\begin{eqnarray}\label{PhiL}
\Phi_{L}(r)=\frac{n-1}{2}\sqrt{\mathcal{B}(z-1)}\,\, \rm ln \bigg[\frac{r}{b}\bigg],
\end{eqnarray}
\begin{eqnarray}\label{Lambda1}
\Lambda_{L}=-\frac{\mathcal{B}}{2l^{2}}[z^2+(2n-3)z]-\frac{n-1}{2l^2}\bigg[n-2-\frac{n-4}{l^2}\hat{\alpha}_{2}+\frac{n-6}{l^4}\hat{\alpha}_{3}-\frac{n-8}{l^6}\hat{\alpha}_{4}\bigg],
\end{eqnarray}
\begin{eqnarray}\label{q1}
q^{2}_{L}=\frac{\mathcal{B}}{2}(z-1)(n+z-1)b^{2n-2},
\end{eqnarray}
\begin{eqnarray}\label{lambda1}
\lambda_{L}=\frac{n-1}{\sqrt{\mathcal{B}(z-1)}},
\end{eqnarray}
where $b$ shows a constant of integration and $\mathcal{B}=1-\frac{2\hat{\alpha}_{2}}{l^2}+\frac{3\hat{\alpha}_{3}}{l^4}-\frac{4\hat{\alpha}_{4}}{l^6}$.
\section{field equations}\label{eauation}
In this part, we obtain the field equations of the rotating Lifshitz dilaton black brane in the presence of the quartic quasitopological gravity. For simplification, we change the metric \eqref{metric2} into the following one
\begin{eqnarray}\label{met2}
ds^2=-e^{2A(r)}(\Xi dt-ad\phi)^2+e^{2C(r)}dr^2+\frac{e^{2B(r)}}{l^2}(adt-\Xi l^2 d\phi)^2+l^2e^{2B(r)}\sum_{i=1}^{n-2}d \vec{X}_{i}^2,
\end{eqnarray}
where we have used of the transformations
\begin{eqnarray}\label{trans2}
A(r)&=&\frac{1}{2} \mathrm{ln}\bigg[\frac{r^{2z}}{l^{2z}}f(r)\bigg],\nonumber\\
C(r)&=&-\frac{1}{2} \mathrm{ln}\bigg[\frac{r^{2}}{l^{2}}g(r)\bigg],\nonumber\\
B(r)&=&\mathrm{ln}\bigg[\frac{r}{l}\bigg].\nonumber\\
\end{eqnarray}
Now if we replace the definitions \eqref{Lag}- \eqref{pot}, \eqref{ans} and \eqref{met2} in the action \eqref{Action} and then integrate by parts, we get to the Lagrangian
\begin{eqnarray}\label{Lag1}
\mathcal{L}&=&l^{n-1}e^{(n-1)B}\bigg(-2\Lambda e^{A+C}+2q^2 K^{'2}e^{-A-C-4\lambda \Phi/(n-1)+2K}+e^{A-C}\big[2(n-1) A^{'}B^{'}+(n-1)(n-2)B^{'2}\nonumber\\
&&-\frac{4}{n-1}\Phi^{'2}\big]-\frac{4(n-1)}{3}\hat{\alpha}_{2}e^{A-3C}\bigg[A^{'}B^{'3}+\frac{(n-4)}{4}B^{'4}\bigg]\nonumber\\
&&+\frac{6(n-1)}{5}\hat{\alpha}_{3}e^{A-5C}\bigg[A^{'}B^{'5}+\frac{(n-6)}{6}B^{'6}\bigg]-\frac{8(n-1)}{7}\hat{\alpha}_{4}e^{A-7C}\bigg[A^{'}B^{'7}+\frac{(n-8)}{8}B^{'8}\bigg]\bigg).
\end{eqnarray}
By varying the action \eqref{Lag1} with respect to $A(r)$, $B(r)$, $C(r)$, $\Phi(r)$ and $K(r)$, the field equations are obtained as bellow
\begin{eqnarray}\label{E1}
E_{1}&=&\frac{\Lambda}{n-1}e^{A+C+(n-1)B}+\frac{q^2}{n-1}K^{'2}e^{-A-C+(n-1)B+2K-4\lambda\Phi/(n-1)}+[B^{''}+\frac{n}{2}B^{'2}-B^{'}C^{'}+\frac{2}{(n-1)^2}\Phi^{'2}]e^{A-C+(n-1)B}\nonumber\\
&&-\frac{n}{2}\hat{\alpha}_{2} [\frac{4}{n}B^{''}B^{'2}+B^{'4}-\frac{4}{n}B^{'3}C^{'}]e^{A-3C+(n-1)B}+\frac{n}{2}\hat{\alpha}_{3} [\frac{6}{n}B^{'4}B^{''}-\frac{6}{n}B^{'5}C^{'}+B^{'6}]e^{A+(n-1)B-5C}\nonumber\\
&&-\frac{n}{2}\hat{\alpha}_{4} [\frac{8}{n}B^{'6}B^{''}-\frac{8}{n}B^{'7}C^{'}+B^{'8}]e^{A+(n-1)B-7C}=0,
\end{eqnarray}
\begin{eqnarray}\label{E2}
E_{2}&=&e^{A+C+(n-1)B}\Lambda-q^2 K^{'2}e^{-A-C+(n-1)B+2K-4\lambda\Phi/(n-1)}\nonumber\\&&-[-(n-2)B^{''}-A^{''}-\frac{(n-1)(n-2)}{2}B^{'2}-A^{'2}+(n-2)B^{'}(C^{'}-A^{'})+A^{'}C^{'}-\frac{2}{n-1}\Phi^{'2}]e^{A-C+(n-1)B}\nonumber\\
&&-2\hat{\alpha}_{2} B^{'}
\bigg\{2A^{'}B^{''}+B^{'}\bigg[B^{'}\bigg(\frac{(n-1)(n-4)}{4}B^{'}+(n-2)A^{'}-(n-4)C^{'}\bigg)+A^{'}(A^{'}-3C^{'})+(n-4)B^{''}+A^{''}\bigg]\bigg\}  \nonumber\\
&&\times e^{A-3C+(n-1)B}+
\nonumber\\
&&+3\hat{\alpha}_{3} B^{'3}\bigg\{4A^{'}B^{''}+B^{'}\bigg[B^{'}\bigg(\frac{(n-1)(n-6)}{6}B^{'}+(n-2)A^{'}-(n-6)C^{'}\bigg)+A^{'}(A^{'}-5C^{'})+(n-6)B^{''}+A^{''}\bigg]\bigg\}\nonumber\\
&&\times e^{A+(n-1)B-5C}+\nonumber\\
&&-4\hat{\alpha}_{4} B^{'5}\bigg\{6A^{'}B^{''}+B^{'}\bigg[B^{'}\bigg(\frac{(n-1)(n-8)}{8}B^{'}+(n-2)A^{'}-(n-8)C^{'}\bigg)+A^{'}(A^{'}-7C^{'})+(n-8)B^{''}+A^{''}\bigg]\bigg\}\nonumber\\
&&\times e^{A+(n-1)B-7C}=0,
\end{eqnarray}
\begin{eqnarray}\label{E3}
E_{3}&=& \frac{\Lambda}{n-1}e^{A+C+(n-1)B}+\frac{q^2}{n-1} K^{'2}e^{-A-C+(n-1)B+2K-4\lambda\Phi/(n-1)}+[A^{'}B^{'}+\frac{n-2}{2}B^{'2}-\frac{2}{(n-1)^2}\Phi^{'2}]e^{A-C+(n-1)B}\nonumber\\
&&-2\hat{\alpha}_{2} B^{'3} [A^{'}+\frac{n-4}{4}B^{'}] e^{A-3C+(n-1)B}
+3\hat{\alpha}_{3} B^{'5} [A^{'}+\frac{n-6}{6}B^{'}]e^{A+(n-1)B-5C}\nonumber\\
&&-4\hat{\alpha}_{4} B^{'7} [A^{'}+\frac{n-8}{8}B^{'}]e^{A+(n-1)B-7C}=0,
\end{eqnarray}
\begin{eqnarray}\label{E4}
E_{4}&=& -\frac{4}{(n-1)^2}\{[\Phi^{''}+\Phi^{'}(A^{'}-C^{'}+(n-1)B^{'})]e^{A-C+(n-1)B}-\lambda q^2 K^{'2}e^{-A-C+(n-1)B+2K-4\lambda\Phi/(n-1)}\}=0,
\end{eqnarray}
\begin{eqnarray}\label{E5}
E_{5}&=& \frac{2q^2}{(n-1)}\bigg\{K^{''}+K^{'2}+K^{'}\bigg[-\frac{4}{n-1}\lambda\Phi^{'}-A^{'}-C^{'}+(n-1)B^{'}\bigg]\bigg\}e^{-A-C+(n-1)B+2K-4\lambda\Phi/(n-1)}=0,
\end{eqnarray}
where a prime denotes the derivative with respect to the $r$ coordinate. It should be noted that the field equations and so the solutions of the rotating case are the same as the static ones, since the rotating metric \eqref{metric2} and the static one \eqref{metric1} are locally the same.\\
The obtained field equations are not solved analytically and we can just get to the bellow solution for the final equation $E_{5}$ as 
\begin{eqnarray}\label{maxwell2}
\frac{d}{dr} k(r)=r^{z-n}\sqrt{\frac{f(r)}{g(r)}} e^{4\lambda\Phi(r)/(n-1)},
\end{eqnarray}
where we have used of the transformations \eqref{trans1} and \eqref{trans2}. 
If we combine the field equations, we can get to a constant where can help to obtain the Smarr-type formula in the thermodynamics without having any exact black brane solutions. So this constant along the radial coordinate is obtained as bellow
\begin{eqnarray}
-2(n-1)l^{n-1}\big(E_{1}-\frac{E_{2}}{n-1}-\frac{n-1}{2\lambda}E_{4}\big)&=&\mathcal{C}^{'}=0,
\end{eqnarray}
where  
\begin{eqnarray}\label{Ctot}
\mathcal{C}&=&2l^{n-1}[e^{A-C+(n-1)B}(A^{'}-B^{'})+2\hat{\alpha}_{2} e^{A-3C+(n-1)B}(B^{'3}-A^{'}B^{'2})-3\hat{\alpha}_{3} e^{A+(n-1)B-5C}(B^{'5}-A^{'}B^{'4})\nonumber\\
&&+4\hat{\alpha}_{4} e^{A+(n-1)B-7C}(B^{'7}-A^{'}B^{'6})-\frac{2}{\lambda}e^{A-C+(n-1)B}\Phi^{'}].
\end{eqnarray}
Now, if we use the relations \eqref{trans1}, \eqref{trans2} and \eqref{maxwell2} in Eq. \eqref{Ctot}, then we attin the total constant as  
\begin{eqnarray}\label{Ctotal}
\mathcal{C}=\frac{r^{z+n}}{l^{z+1}} \bigg[\mathcal{B}\bigg(f^{'}\sqrt{\frac{g}{f}}+\frac{2(z-1)}{r}\sqrt{fg}\bigg)-\frac{4\Phi^{'}}{\lambda}\sqrt{fg}\bigg].
\end{eqnarray}
As it not possible to obtain the exact solutions for the quartic quasitopological rotating Lifshitz dilaton black brane, so we can probe their asymptotic behaviors in a near-horizon and infinity regimes in the later section. 
\section{near-horizon and infinity behaviors}\label{asy}
At first, we look for the well-behaved black hole solutions in a near-horizon regime. The functions expanstions near the horizon are 
\begin{eqnarray}\label{basthorizon}
f(r)&=& f_{1}\{(r-r_{+})+f_{2} (r-r_{+})^2+f_{3} (r-r_{+})^3+...\}\nonumber\\
g(r)&=& g_{1}(r-r_{+})+g_{2} (r-r_{+})^2+g_{3} (r-r_{+})^3+...\nonumber\\
\Phi(r)&=& \Phi_{+}+\Phi_{1}(r-r_{+})+\Phi_{2} (r-r_{+})^2+\Phi_{3} (r-r_{+})^3... ,
\end{eqnarray}
where $r_{+}$ is the event horizon radius. The coefficients $f_{i}$, $g_{i}$ and $\Phi_{i}$'s are defined by substituting the expansions \eqref{basthorizon}, the metric \eqref{met2} and the constraints \eqref{Lambda1}-\eqref{lambda1} in the equations of motion \eqref{E1}-\eqref{E4} and solving for various equations. For a nonextreme rotating black brane, the functions $f(r)$ and $g(r)$ go to zero linearly, however the dilaton field has a nonzero value at the horizon and so $\Phi_{+}\neq 0$. 
Using the expansions \eqref{basthorizon} in the relation \eqref{Ctotal}, we have the constant at the horizon $r=r_{+}$ as
\begin{eqnarray}\label{Cthorizon}
\mathcal{C}=\frac{r_{+}^{z+n}}{l^{z+1}}\sqrt{f_{1} g_{1}},
\end{eqnarray}
where it will appeared in the next section. We would also like to probe the rotating quasitopological Lifshitz dilaton black brane solutions at the infinity. So we use the bellow straightforward perturbation theory
\begin{eqnarray}\label{exinfinity}
f(r)&=& 1+\epsilon f_{1}(r)\nonumber\\
g(r)&=& 1+\epsilon g_{1}(r)\nonumber\\
\Phi(r)&=& \Phi_{L}(r)+\epsilon \Phi_{1}(r),
\end{eqnarray}
where $\epsilon$ is an infinitesimal parameter. Using the expansions \eqref{exinfinity}, the conditions \eqref{PhiL}-\eqref{lambda1} and the transformations \eqref{trans1} and \eqref{trans2} in the equations of motion \eqref{E1}-\eqref{E4}, we can find the field equations up to the first order of $\epsilon$ as the following
\begin{eqnarray}\label{Eas1}
&&(n-1)\sqrt{\mathcal{B}}[r g_{1}^{'}+(n+z-1)g_{1}]+4\sqrt{z-1}\,[r\Phi_{1}^{'}+(n+z-1)\Phi_{1}]=0,
\end{eqnarray}
\begin{eqnarray}
&&\sqrt{\mathcal{B}}[(n+z-2)rg_{1}^{'}+(n+z-1)(n+2z-3)g_{1}+r^2f_{1}^{''}+(n+2z-1)rf_{1}^{'}]\nonumber\\
&&+\frac{4}{\sqrt{\mathcal{B}}}(z-1)(-\frac{\hat{\alpha}_{2}}{l^2}+\frac{3\hat{\alpha}_{3}}{l^4}-\frac{6\hat{\alpha}_{4}}{l^6})[rg_{1}^{'}+(n+z-1)g_{1}]+4\sqrt{z-1}[r\Phi_{1}^{'}-(n+z-1)\Phi_{1}]=0,
\end{eqnarray}
\begin{eqnarray}
&&(n-1)\sqrt{\mathcal{B}}[(n+z-1)g_{1}+rf_{1}^{'}] +\frac{4}{\sqrt{\mathcal{B}} }(z-1)(n-1)\big[-\frac{\hat{\alpha}_{2}}{l^2}+\frac{3\hat{\alpha}_{3}}{l^4}-\frac{6\hat{\alpha}_{4}}{l^6}\big]g_{1}\nonumber\\
&&+4\sqrt{z-1}[(n+z-1)\Phi_{1}-r\Phi_{1}^{'}]=0,
\end{eqnarray}
\begin{eqnarray}\label{Eas4}
&&(n-1)\sqrt{(z-1)\mathcal{B}}[r(g_{1}^{'}+f_{1}^{'})+2(n+z-1)g_{1}]+4[r^2\Phi_{1}^{''}+r(n+z)\Phi_{1}^{'}-2(n-1)(n+z-1)\Phi_{1}]=0.
\end{eqnarray}
Now if we solve the above equations, we can achieve the functions at the infinity as 
\begin{eqnarray}\label{finfi}
f_{1}(r)&=&-\frac{\mathcal{F}_{1}}{r^{n+z-1}}-\frac{\mathcal{F}_{2}}{r^{(n+z+\gamma-1)/2}},\nonumber\\
g_{1}(r)&=&-\frac{\mathcal{W}_{1} \mathcal{F}_{1}}{r^{n+z-1}}-\frac{\mathcal{W}_{2} \mathcal{F}_{2}}{r^{(n+z+\gamma-1)/2}},\nonumber\\
\Phi_{1}(r)&=&-\frac{\mathcal{W}_{3} \mathcal{F}_{1}}{r^{n+z-1}}-\frac{\mathcal{W}_{4} \mathcal{F}_{2}}{r^{(n+z+\gamma-1)/2}},
\end{eqnarray}
where 
\begin{eqnarray}
\mathcal{W}_{1}&=&(n+z-1)(n+z-2)\mathcal{B}\mathcal{T}^{-1},\nonumber\\
\mathcal{W}_{2}&=&-(n+z+\gamma-1)\mathcal{B}\mathcal{Z}^{-1},\nonumber\\
\mathcal{W}_{3}&=&\frac{n-1}{2}(z-1)^{3/2}\bigg(\frac{\hat{\alpha}_{2}} {l^2}-\frac{3\hat{\alpha}_{3}}{l^4}+\frac{6\hat{\alpha}_{4}}{l^6}\bigg)\mathcal{B}^{1/2}\,\mathcal{T}^{-1},\nonumber\\
\mathcal{W}_{4}&=&\frac{n-1}{4}(n+z+\gamma-1)(z-1)^{-1/2}\mathcal{B}^{3/2}\mathcal{Z}^{-1},\nonumber\\
\mathcal{T}&=&(n+z-1)(n+z-2)\mathcal{B}-4(n-1)(z-1)\bigg[\frac{\hat{\alpha}_{2}}{l^2}-\frac{3\hat{\alpha}_{3}}{l^4}+\frac{6\hat{\alpha}_{4}}{l^6}\bigg],\nonumber\\
\mathcal{Z}&=&(n+z+\gamma-1)\mathcal{B}+8(z-1)\bigg[\frac{\hat{\alpha}_{2}}{l^2}-\frac{3\hat{\alpha}_{3}}{l^4}+\frac{6\hat{\alpha}_{4}}{l^6}\bigg],\nonumber\\
\gamma^2&=&(n+z-1)(9n+9z-17)+16(z-1)^2\mathcal{B}^{-1}\bigg[\frac{\hat{\alpha}_{2}}{l^2}-\frac{3\hat{\alpha}_{3}}{l^4}+\frac{6\hat{\alpha}_{4}}{l^6}\bigg].\nonumber\\
\end{eqnarray}
It is clear from Eq. \eqref{finfi} that in the limit $r\rightarrow\infty$, the functions $f_{1}(r)$, $g_{1}(r)$ and $\Phi_{1}(r)$ go to zero and so $f(r)$ and $g(r) \rightarrow 1$ and $\Phi(r)\rightarrow \Phi_{L}(r)$. If we use the functions \eqref{finfi} in Eq. \eqref{Ctotal}, then the constant at infinity reduces to
\begin{eqnarray}\label{mathcalC}
\mathcal{C}=\frac{(n+z-1)\mathcal{B}[(n+z-2)(n+z-1)\mathcal{B}+2(z-1)^2(\frac{\hat{\alpha}_{2}}{l^2}-\frac{3\hat{\alpha}_{3}}{l^4}+\frac{6\hat{\alpha}_{4}}{l^6})]}{\mathcal{T}l^{z+1}}\mathcal{F}_{1}.
\end{eqnarray}
\section{Thermodynamics}\label{thermo}
Our purpose in this section is to study thermodynamics of the rotating Lifshitz dilaton black brane in the quartic quasitopological gravity. At first, we make a finite well-defined action and then obtain the conserved and thermodynamic quantities. We also investigate thermal stability of the mentioned rotating black brane.\\
In the case of a manifold with boundary, the action \eqref{Action} is neither well-defined nor finite. This makes us to add some surface terms and counterterms to this action in order to have a well-defined variational principle and a finite action, respectively. For the spacetimes with flat boundaries, we choose the total surface term as  
\begin{eqnarray}
I_{s}&=&\frac{1}{8\pi}\int_{\partial\mathcal{M}}d^{n} x \sqrt{-h}\times\nonumber\\
&&\bigg[K+\frac{2\hat{\alpha}_{2}}{(n-2)(n-3)} J+\frac{3\hat{\alpha}_{3}}{5n(n-2)(n-1)^2(n-5)}H+\frac{2\hat{\alpha}_{4}}{7n(n-1)(n-2)(n-7)(n^2-3n+3)}Q\bigg],
\end{eqnarray}
where $h$ is the determinant of the induced metric $h_{\mu\nu}$ on the boundary $\partial{\mathcal{M}}$. The Gibbons and Hawking, Gauss-Bonnet and cubic quasitopological surface terms are defined by $K$, $J$ and $H$ \cite{Myers00,SCQ} where are the trace of the bellow tensors 
\begin{eqnarray}
K&=&K_{a b}h^{a b}\nonumber\\
J_{ab}&=&\frac{1}{3}(2KK_{a\gamma}K^{\gamma}_{b}+K_{\gamma\delta}K^{\gamma\delta}K_{ab}-2K_{a\gamma}K^{\gamma\delta}K_{\delta b}-K^2K_{ab}),\nonumber\\
H_{ab}&=&nK^4 K_{ab}-nK^2 K_{ab}K_{cd}K^{cd}+(n-2)K^3 K_{a}^{c}K_{c b}+(n-1)(3n-2)K_{ac}K^{c}_{b}K_{m d}K^{d}_{e}K^{em}\nonumber\\
&&-3(n-1)(n-2)K_{c d}K^{cd}K_{a e}K^{e m}K_{m b}-n(5n-6)K K_{c d}K^{c d}K_{am}K^{m}_{b}\nonumber\\
&&+(n-1)(5n-6)K K_{a r}K^{c}_{b}K^{r d}K_{c d},
\end{eqnarray}
that $K_{\mu\nu}$ is the extrinsic curvature. Surface term of the quartic quasitopological gravity has been described in Ref. \cite{SQQ},
where we could have achieved the related tensor as bellow 
\begin{eqnarray}\label{QS}
Q_{ab}&=&\frac{\beta_{1}}{3}\{(3-2a_{1})K^2K_{a b}K^{m n}K_{m c}K_{n d}K^{c d}+2a_{1} K^3 K_{a} ^{n}K_{b c}K_{n d}K^{c d}\}+\beta_{2}\{(4-a_{2}-2a_{3})K ^2K_{b}^{n}K_{a n}K^{c d}K^{e}_{c}K_{d e}\nonumber\\
&&+a_{2}K K_{a b}K^{m n}K_{m n}K^{c d}K_{c}^{e}K_{d e}+2a_{3} K^2 K^{m n}K_{m n}K_{a}^{d}K_{d e}K^{e}_{b}\}+\beta_{3}\{KK_{a b}K^{m n}K_{m c}K_{n d}K^{c e}K_{e}^{d}\nonumber\\
&&+3K^2K_{b}^{n}K_{a c}K_{n d}K^{c e}K_{e}^{d}\}+\frac{\beta_{4}}{5}\{(2-a_{4})K_{a b}K^{m n}K_{m n}K^{c d}K_{c}^{e}K_{d}^{f}K_{e f}+a_{4}K K_{b}^{n}K_{a n}K^{c d}K_{c}^{e}K_{d}^{f}K_{e f}\nonumber\\
&&+3KK^{m n}K_{m n}K_{b}^{d}K_{a}^{e}K_{d}^{f}K_{e f}\}+\frac{\beta_{5}}{5}\{K_{a b}K^{m n}K_{m}^{c}K_{n c}K^{d e}K_{d}^{f}K_{e f}+4K K^{m n}K_{m}^{c}K_{n c}K_{b}^{e}K_{a}^{f}K_{e f}\}\nonumber\\
&&+\frac{\beta_{6}}{2}\{(4-3a_{5})K_{a b }K^{m n}K_{m c}K_{n d}K^{c e}K^{d f}K_{e f}+3a_{5}KK_{b}^{n}K_{a c}K_{n d}K^{c e}K^{d f}K_{e f}\}\nonumber\\
&&+\beta_{7}\{K^{m n}K_{m b}K_{a n}K^{d e}K_{d f}K_{e g}K^{f g}+K^{m n}K_{m}^{c}K_{n c}K_{b}^{e}K_{a f}K_{e g}K^{f g}\}
\end{eqnarray}
that the coefficients $\beta_{i}$'s and $a_{i}$'s have been written in App. \eqref{app2}.
We can have a finite action by combining a reference term in
the spacetime that has been introduced through the substraction method of Brown and York \cite{Brown}. 
By this method, we recommend the counterterm action for the rotating Lifshitz dilaton black brane in the quartic quasitopological gravity by
\begin{eqnarray}
I_{c}=-\frac{1}{8\pi} \int_{\partial \mathcal{M}} d^{n} x \sqrt{-h} \bigg[\frac{(n-1)(l^6-2\hat{\alpha}_{2}l^4+3\hat{\alpha}_{3}l^2-4\hat{\alpha}_{4})}{l^7}+\frac{2q}{lb^{n-1}}e^{-2\lambda\Phi/(n-1)}\sqrt{-A_{\gamma}A^{\gamma}}\bigg],
\end{eqnarray}
where it is a functional of the boundary curvature invariants \cite{Hennin,Hyun}. Now if we vary of the total action $I=I_{b}+I_{s}+I_{c}$ about the solutions, we get to 
\begin{eqnarray}
\delta I_{\mathrm{tot}}=\int d^{n}x (S_{ab}\delta h^{ab}+S_{a}^{L}\delta U^{a}),
\end{eqnarray}
where 
\begin{eqnarray}\label{Sab}
S_{ab}=\frac{\sqrt{-h}}{16\pi}\{\Pi_{ab}+\frac{2q}{lb^{n-1}}e^{-2\lambda\Phi/(n-1)}(-A_{\gamma}A^{\gamma})^{-1/2}(A_{a}A_{b}-A_{\gamma}A^{\gamma}h_{ab})\},
\end{eqnarray}
\begin{eqnarray}
S_{b}^{L}=-\frac{\sqrt{-h}}{16\pi}e^{-4\lambda\Phi/(n-1)}\{4n^{a}F_{a b}-\frac{4q}{lb^{n-1}}e^{2\lambda\Phi/(n-1)}(-A_{\gamma}A^{\gamma})^{-1/2}A_{b}\},
\end{eqnarray}
and 
\begin{eqnarray}
\Pi_{ab}&=&K_{ab}-Kh_{ab}+\frac{2\hat{\alpha}_{2}}{(n-2)(n-3)}[3J_{ab}-Jh_{ab}]+\frac{3\hat{\alpha}_{3}}{5n (n-2)(n-5)(n-1)^2}[5H_{ab}-H h_{ab}]\nonumber\\
&&+\frac{2\hat{\alpha}_{4}}{7n (n-1)(n-2)(n-7)(n^2-3n+3)}[7Q_{ab}-Q h_{ab}]+\frac{(n-1)[l^6-2\hat{\alpha}_{2}l^4+3\hat{\alpha}_{3}l^2-4\hat{\alpha}_{4}]}{l^7}h_{ab}.
\end{eqnarray}
As we said in the introduction, the dual field theory for an asymptotically Lifshitz spacetime is nonrelativistic and so it will not possess a covariant relativistic stress tensor. One can define a stress
tensor complex that includes the energy density $\mathcal{E}$, energy flux $\mathcal{E}_{i}$, momentum density $\mathcal{P}_{i}$,
and spatial stress tensor $\mathcal{P}_{ij}$ with the definitions \cite{Ross} 
\begin{eqnarray}\label{consqua}
\mathcal{M}=2S^{t} _{t}-S^{t}A_{t},\,\,\,\,\,\,\, \mathcal{M}^{i}=2S^{i}_{t}-S^{i}A_{t},
\end{eqnarray}
and
\begin{eqnarray}\label{aa}
\mathcal{P}_{i}=-2S^{t}_{i}+S^{t}A_{i},\,\,\,\,\,\, \mathcal{P}^{j}_{i}=-2S^{j}_{i}+S^{j}A_{i}.
\end{eqnarray}
The above quantities establish the bellow conservation equations  \cite{Ross}
\begin{eqnarray} 
\partial_{t} \mathcal{M}+\partial_{i} \mathcal{M}^{i}=0,\,\,\,\,\,\,\,\partial_{t} \mathcal{P}_{j}+\partial_{i} \mathcal{P}^{i}_{j}=0.
\end{eqnarray}
Using Eqs. \eqref{consqua} and \eqref{aa} and the expansions \eqref{exinfinity}, we can get to the energy and angular momentum densities of the rotating quartic quasitopological Lifshitz dilaton black brane as
\begin{eqnarray}\label{mass}
\mathcal{M}=\frac{(n+z-1)\Xi^2-z}{16\pi(n+z-1)}\mathcal{C},
\end{eqnarray}
\begin{eqnarray}\label{angular}
J=\frac{a\Xi}{16\pi} \mathcal{C},
\end{eqnarray}
where $\mathcal{C}$ is the constant at the infinity in Eq. \eqref{mathcalC}. For $\Xi=1$, they result to 
\begin{eqnarray}
\mathcal{M}=\frac{n-1}{16\pi(n+z-1)}\mathcal{C},\,\,\, J=0,
\end{eqnarray}
for the static quartic quasitopological Lifshitz dilaton black brane. 
The killing vector of a rotating black brane is defined by
\begin{eqnarray}
\chi=\partial_{t}+\Omega \partial_{\phi},
\end{eqnarray}
where $\Omega$ demonstrates the angular velocity of the Killing horizon by 
\begin{eqnarray}\label{Omega}
\Omega=-\bigg[\frac{g_{t\phi}}{g_{\phi\phi}}\bigg]_{r=r_{+}}=\frac{a}{\Xi l^2}.
\end{eqnarray}
We can also calculate the entropy of the rotating quartic quasitopological Lifshitz dilaton black brane with a flat horizon as the following \cite{Wald}
\begin{eqnarray}\label{entropy}
S=\frac{1}{4}\Xi r_{+}^{n-1}.
\end{eqnarray}
The temperature of the mentioned rotating black brane at the event horizon is also attained by  
\begin{eqnarray}\label{temp}
T=\frac{1}{2\pi}\sqrt{-\frac{1}{2}\nabla_{b}\chi_{a}\nabla^{b}\chi^{a}}\mid_{r=r_{+}}=\bigg(\frac{r^{z+1} f^{'}g^{'}}{4\pi l^{z+1}\Xi}\bigg)_{r=r_{+}}=\frac{r_{+}^{z+1}}{4\pi l^{z+1}\Xi}\sqrt{f_{1}g_{1}},
\end{eqnarray}
where we have used of the near horizon expansions \eqref{basthorizon} in order to gain the last term. 
For $\Xi=1$, the entropy and temperature of the static quartic quasitopological Lifshitz dilaton black brane are appeared
\begin{eqnarray}
S=\frac{1}{4}r_{+}^{n-1},\,\,\,\, T=\frac{r_{+}^{z+1}}{4\pi l^{z+1}}\sqrt{f_{1}g_{1}}.
\end{eqnarray}
According to the expansions in \eqref{basthorizon}, $f_{1},g_{1}\propto\frac{1}{r_{+}}$ and so we can conclude that
\begin{eqnarray}\label{Temp}
T =\frac{\eta}{4\pi \Xi}r_{+}^{z},
\end{eqnarray}
where $\eta$ represents a proportionality constant. Using Eqs. \eqref{Cthorizon}, \eqref{entropy} and \eqref{temp}, we can get to the bellow relation
\begin{eqnarray}\label{CTS}
\mathcal{C}=16\pi T S.
\end{eqnarray}
Now if we combine the relations \eqref{mass}, \eqref{angular} and \eqref{CTS}, then we can achieve the bellow relation 
\begin{eqnarray}
\mathcal{M}=\frac{n-1}{n+z-1}TS+\Omega J.
\end{eqnarray} 
Dividing the mass \eqref{mass} by the angular momentum \eqref{angular}, we obtain a Smarr-type formula such as 
\begin{eqnarray}
\mathcal{M}(S,J)=\frac{(n+z-1)Z-z}{(n+z-1)l\sqrt{Z(Z-1)}}J,
\end{eqnarray}
where $Z=\Xi^2$ depends on $S$ and $J$ through the following equation
\begin{eqnarray}
16\pi J-\eta l \bigg(\frac{4S}{\Xi}\bigg)^{(n+z-1)/(n-1)}\sqrt{Z(Z-1)}=0.
\end{eqnarray} 
If we consider the parameters $S$ and $J$ as a complete set of the extensive parameters for
the energy density $\mathcal{M}(S,J)$, then the intensive parameters conjugate to $S$ and $J$ are defined as bellow 
\begin{eqnarray}\label{intensive}
T=\bigg(\frac{\partial \mathcal{M}}{\partial S}\bigg)_{J},\,\,\,\,\,\Omega=\bigg(\frac{\partial \mathcal{M}}{\partial J}\bigg)_{S}.
\end{eqnarray}
By some calculations, we can show that the intensive parameters evaluated in Eq. \eqref{intensive} are the same as the ones in Eqs.\eqref{Omega} and \eqref{Temp}. Thus we can deduce that the rotating quartic quasitopological Lifshitz dilaton black brane obeys the first law of the thermodynamics
\begin{eqnarray}
d\mathcal{M}=TdS+\Omega dJ.
\end{eqnarray}
We also tend to probe thermal stability of the rotating quartic quasitopological Lifshitz dilaton black brane. If we consider the energy density as a function of the extensive parameters $S$ and $J$, then the solutions are thermally stable if the bellow conditions are satisfied
\begin{eqnarray}
T&>&0,\nonumber\\
C=T(\partial T/\partial S)&>&0,\nonumber\\
Det(H)=\bigg(\frac{\partial^{2} \mathcal{M}}{\partial S^{2}}\bigg)\bigg(\frac{\partial^{2} \mathcal{M}}{\partial J^{2}}\bigg)-\bigg(\frac{\partial^{2} \mathcal{M}}{\partial S\partial J}\bigg)^2&>&0.
\end{eqnarray} 
where $C$ denotes the heat capacity and $Det(H)$ is the determinant of the energy Hessian matrix with the definition 
\begin{eqnarray}
H=\left[
\begin{array}{ccc}
\frac{\partial^2 M}{\partial S^2} & \frac{\partial^2 M}{\partial S\partial J} \\
\frac{\partial^2 M}{\partial S \partial J}& \frac{\partial^2 M}{\partial J^2} 
\end{array} \right].
\end{eqnarray}
It's a matter of calculation to show that $C$ and $Det(H)$ are obtained as bellow
\begin{eqnarray}\label{cap}
C=\frac{\eta^2[(n+3z-1)Z-n-2z+1]}{64\pi^2 S^2\sqrt{Z}[(n-z-1)Z+z]}r_{+}^{2z+n-1},
\end{eqnarray}
\begin{eqnarray}\label{det}
Det(H)&=&-\frac{(Z-1)(n+z-1)^2}{ZS ^2 l^2[(n-z-1)Z+z]^2}-\frac{[(n+3z-1)Z-n-2z+1]}{4\pi Zl^2S^2(Z-1)(n+z-1)[(n-z-1)Z+z]^2}\nonumber\\
&&\times\bigg[4\pi(n-1)[(n-z-1)Z+z]+l\eta(n+z-1)\sqrt{Z-1}(n-1)(2Z-1)\bigg(\frac{4S}{\sqrt{Z}}\bigg)^{\frac{z}{n-1}}\bigg].
\end{eqnarray}
The temperature \eqref{Temp} is positive for a positive value of $\eta$. We have also defined $Z=\Xi^2=1+a^2/l^2$ in the previous sections and as the parameter $a$ has a nonzero value, we have the result $Z>1$. We can deduce that for $z>1$, $Z>1$, $n>4$ and $\eta>0$, the heat capacity $C$ and $Det(H)$ are respectively positive and negative for any other parameters. By this, we can conclude that the rotating quartic quasitopological Lifshitz dilaton black brane is not thermally stable. 
\section{RESULTS}\label{result}
In this paper, we could reach the quartic quasitopological Lifshitz dilaton solutions for the static and rotating black branes with the rotation parameter $a$. The obtained solutions have the Lifshitz
symmetry $t\rightarrow \zeta^{z}t$ and $\vec{X}\rightarrow \zeta \vec{X}$ at $r$ infinity boundary where $z$ represents the dynamical critical
exponent. Based on the AdS/CFT correspondence, quasitopological gravity can provide a
broader class of four (and higher)-dimensional CFT’s whose central charges are in accordance with the coupling parameters in the gravitational spacetime. Quartic quasitopological gravity with fourth-order curvature tensors can lead to the field equations in all dimensions $n\geq4$, except for $n=7$. This is while that the $k$-th order LoveLock theory is effective only for $n\geq 2k$.\\
In this paper, we first obtained some constraints on $\Phi(r)$, $\Lambda$, $q$ and $\lambda$ to ensure the existence of the asymptotic Lifshitz solutions. Then we varied the action with respect to the functions and attained five field equations where were independent of the rotation. So we deduced that the solutions of the static and rotating quartic quasitopological Lifshitz dilaton black branes are the same, since their related metrics are mapped onto each other locally, not globally. Using the field equations, we could gain a constant $\mathcal{C}$ along the radial coordinate $r$ which was usable in the thermodynamics. It was not possible to find the solutions exactly, so we probed them at the near horizon and infinty limits and earned the constant $\mathcal{C}$ at these limits.\\
As the related action was neither well-defined nor finite, so we used some surface terms and counterterms in order to get to a finite well-defined action. We obtained the tensor of the quartic quasitopological surface term and then defined a stress tensor complex in order to calculate the energy and the angular momentum densities which were dependent on the constant $\mathcal{C}$. We also got to the thermodynamic quantities such as temperature $T$, entropy $S$ and angular velocity $\Omega$ where $\mathcal{C}$ was related to the product of $T$ and $S$. Removing the constant $C$, we could achieve a relation between the thermodynamic quantities an then reach a Smarr-type formula in which the energy density was versus $S$ and $J$. Our calculations showed that the rotating quartic quasitopological Lifshitz dilaton black brane solutions obey the first law of the thermodynamics. In the absence of the rotation ($a=0$ or $\Xi=1$), the obtained results reduced to the ones for the static case. We also investigated the thermal stability of the rotating quartic quasitopological Lifshitz dilaton black brane. For $z>1$ and $Z=\Xi^2>1$, the temperature $T$ and the heat capacity $C$ are positive but the Hessian matrix determinant is negative. So, this black brane is not thermally stable.\\
Some other studies can be done along this work. One can investigate the critical behavior and Joule-Thomson expansion of the rotating quartic quasitopological Lifshitz dilaton black brane. It is also possible to generalize this study and obtain the solutions of the Lifshitz dilaton black brane in the quintic quasitopological gravity.        
\acknowledgments{We would like to thank Payame Noor University and Jahrom University.}
\section{Appendix}
\subsection{coefficients of the quartic quasitopological gravity}\label{app1}
The coefficients $c_{i}$'s for the quartic quasitopological gravity $\mathcal{L}_{4}$ in Eq. \eqref{Lag} are defined as
\begin{eqnarray}
&&c_{1}=-(n-1)(n^7-3n^6-29n^5+170n^4-349n^3+348n^2-180n+36),\nonumber\\
&&c_{2}=-4(n-3)(2n^6-20n^5+65n^4-81n^3+13n^2+45n-18),\nonumber\\
&&c_{3}=-64(n-1)(3n^2-8n+3)(n^2-3n+3),\nonumber\\
&&c_{4}=-(n^8-6n^7+12n^6-22n^5+114n^4-345n^3+468n^2-270n+54),\nonumber\\
&&c_{5}=16(n-1)(10n^4-51n^3+93n^2-72n+18),\nonumber\\
&&c_{6}=-32(n-1)^2(n-3)^2(3n^2-8n+3),\nonumber\\
&&c_{7}=64(n-2)(n-1)^2(4n^3-18n^2+27n-9),\nonumber\\
&&c_{8}=-96(n-1)(n-2)(2n^4-7n^3+4n^2+6n-3),\nonumber\\
&&c_{9}=16(n-1)^3(2n^4-26n^3+93n^2-117n+36),\nonumber\\
&&c_{10}=n^5-31n^4+168n^3-360n^2+330n-90,\nonumber\\
&&c_{11}=2(6n^6-67n^5+311n^4-742n^3+936n^2-576n+126),\nonumber\\
&&c_{12}=8(7n^5-47n^4+121n^3-141n^2+63n-9),\nonumber\\
&&c_{13}=16n(n-1)(n-2)(n-3)(3n^2-8n+3),\nonumber\\
&&c_{14}=8(n-1)(n^7-4n^6-15n^5+122n^4-287n^3+297n^2-126n+18).\nonumber\\
\end{eqnarray}
\subsection{coefficients of the quartic quasitopological surface terms}\label{app2}
The coefficients $\beta_{i}$'s and $a_{i}$'s in the quartic quasitopological surface term in Eq. \eqref{QS} are 
\begin{eqnarray}
\beta_{1}&=&-4n^2(n-1),\nonumber\\
\beta_{2}&=&-n^3+4n^2-3n-3,\nonumber\\
\beta_{3}&=&3n(n-1)^2\nonumber,\\
\beta_{4}&=&4n(n+2)(n^2-3n+3),\nonumber\\
\beta_{5}&=&2n(n-3)(n^2-n-3),\nonumber\\
\beta_{6}&=&-(n-1)(3n^3-n^2-9n+12),\nonumber\\
\beta_{7}&=&-6,\nonumber\\
\end{eqnarray}
and
\begin{eqnarray}
a_{1}&=&\frac{3(24n^3-79n^2+98n-31)}{40(n-1)(2n^2-5n+5)},\nonumber\\
a_{2}&=&\frac{11n^5-67n^4+148n^3-128n^2+90}{10n^5-65n^4+155n^3-145n^2+75},\nonumber\\
a_{3}&=&\frac{16n^6-125n^5+379n^4-531n^3+261n^2+120n-120}{20n^6-150n^5+440n^4-600n^3+290n^2+150n-150},\nonumber\\
a_{4}&=&\frac{13n^5-54n^4+67n^3+64n^2-225n+165}{8n^5-28n^4+16n^3+88n^2-180n+120},\nonumber\\
a_{5}&=&\frac{2(53n^5-152n^4+30n^3+537n^2-900n+510)}{15(2n^2-5n+5)(3n^3-n^2-9n+12)}.
\end{eqnarray}

%%%%%%%%%%%%%%%%%%%%%%%%%%%%%%%%%%%%%%%%%%%%%%%%%%%%%%%%%%%%%%%%%%

\end{document}